\documentclass[aps,prb,twocolumn,superscriptaddress,showpacs]{revtex4}
\usepackage{amsfonts}
\usepackage{amssymb}
\usepackage{amsmath}
\usepackage{graphicx}

\usepackage{dcolumn}   
\usepackage{bm}        
\usepackage{flafter}

\newcommand{\Tc}{\ensuremath{T_c}}

\begin{document}


\title{Evolution of transport properties of BaFe$_{2-x}$Ru$_x$As$_2$ in a wide range of isovalent Ru substitution}

\date{\today}
\author{M. J. Eom}
\affiliation{Department of Physics, Pohang University of Science and Technology, Pohang 790-784, Korea}
\author{S. W. Na}
\affiliation{Department of Physics, Pohang University of Science and Technology, Pohang 790-784, Korea}
\author{C. Hoch}
\affiliation{Max-Planck-Institut f\"{u}r Festk\"{o}rperforschung, Heisenbergstra$\rm\beta$e 1, D-70569 Stuttgart, Germany}
\author{R. K. Kremer}
\affiliation{Max-Planck-Institut f\"{u}r Festk\"{o}rperforschung, Heisenbergstra$\rm\beta$e 1, D-70569 Stuttgart, Germany}
\author{J. S. Kim}
\email[E-mail:~]{js.kim@postech.ac.kr}
\affiliation{Department of Physics, Pohang University of Science and Technology, Pohang 790-784, Korea}

\begin{abstract}
The effects of isovalent Ru substitution at the Fe sites of BaFe$_{2-x}$Ru$_x$As$_2$ are investigated by measuring resistivity ($\rho$) and Hall coefficient($R_H$) on high-quality single crystals in a wide range of doping (0 $\leq$ $x$ $\leq$ 1.4). Ru substitution weakens the antiferromagnetic (AFM) order, inducing superconductivity for relatively high doping level of 0.4 $\leq$ $x$ $\leq$ 0.9. Near the AFM phase boundary, the transport properties show non-Fermi-liquid-like behavior with a linear-temperature dependence of $\rho$ and a strong temperature dependence of $R_H$ with a sign change. Upon higher doping, however, both $\rho$ and $R_H$ recover conventional Fermi-liquid behavior. Strong doping dependence of $R_H$ together with a small magnetoresistance suggest that the anomalous transport properties can be explained in terms of anisotropic charge carrier scattering due to interband AFM fluctuations rather than a conventional multi-band scenario.

\end{abstract}
\smallskip


\maketitle

Unconventional superconductivity in the proximity of an antiferromagnetic (AFM) phase has been intensively studied on high-$T_c$ cuprates, heavy-fermion superconductors and  the recently-discovered Fe-pnictides.\cite{nonFL:review} In spite of subtle differences in their detailed properties, there is a growing body of evidence that they all exhibit a common phase diagram where inside a dome-shaped regime a superconducting phase appears as the AFM phase is suppressed by an external control parameter, such as doping or pressure. Even though the static AFM order is fully suppressed, AFM fluctuations survive and they can strongly modify the quasi-particle scattering spectrum, leading e. g. to the so-called non-Fermi-liquid-like transport properties. Clarifying the nature of the non-fermi-liquid behavior and understanding its relation to superconductivity are key issues for elucidating the unconventional superconductivity in Fe-pnitides.

In Fe-pnictides, the AFM instability is closely related to the interband nesting between electron- and hole-Fermi surfaces (FS's).\cite{FeAs:mazin:theory,FeAs:eremin:theory,FeAs:Kuroki:theory} Degrading the nesting condition by introducing additional charge carriers or modifying the crystal structure suppress the AFM order and eventually induce superconductivity. So far, various types of chemical substitution have been employed in order to explore the phase diagram of Fe pnictides. In the so-called 122 pnictides, the substitution dependence of the electrical transport properties has been intensively studied \emph{e.g.} for BaFe$_2$As$_2$ with K-\cite{Ba122_K:johrendt:syn}, Co-\cite{Ba122_Co:mandrus:syn,Ba122_Co:fisher:syn,Ba122_Co:hhwen:hall,Ba122_Co:colson:hall} and P-substitution\cite{Ba122_P:matsuda:syn} at the Ba, Fe and As sites, respectively. Deviations from a $T^2$-power-law dependence of the electrical resistivity ($\rho$) or the enhancement of the Hall coefficient ($R_H$) at low temperatures\cite{Ba122_Co:hhwen:hall,Ba122_P:matsuda:syn} have been commonly observed in various Fe-pnictides. These observations are usually considered as experimental indication for non-Fermi-liquid behavior. In Fe-pnictides, however, such deviations have also been ascribed to multi-band transport \cite{Ba122_Co:hhwen:hall,Ba122_Co:colson:hall}, where a conventional description of different types of carriers is sufficient.

In this respect, it is interesting to investigate how the transport properties evolve with isovalent substitution in a wide range. Isovalent substitution does not change the charge carrier density, thus keeping the nature of the exactly-compensated semi-metal. This has been nicely demonstrated in recent studies on P-doped\cite{Ba122_P:matsuda:syn} or Ru-doped\cite{Ba122_Ru:singh:band} BaFe$_2$As$_2$. In particular, Ru substitution at the Fe site is different from other chemical doping in various ways. Firstly, Ru has 4$d$ orbitals, which are spatially more extended than Fe 3$d$ orbitals. This greatly enhances the hybridization with the As $p$ orbitals and thus increases the resulting band width.\cite{Ba122_Ru:singh:band} Secondly, Ru substitution also weakens the electron correlations favoring a nonmagnetic ground state as shown in a recent angle-resolved photoemission spectroscopy (ARPES) study.\cite{Ba122_Ru:colson:ARPES}
Furthermore, chemical pressure by Ru substitution expands the lattice along the $ab$-plane but shrinks it along the $c$-axis.\cite{Ba122_Ru:colson:syn,Ba122_Ru:canfield:syn} This differs from physical pressure or the effect of P-substitution which both lead to shrinkage of the lattice in all directions. Therefore, with Ru substitution the lattice anisotropy is strongly reduced thus modifying the electronic structure more drastically than pressure or P-substitution can do. Investigation of the transport properties of isovalent Ru-substituted BaFe$_2$As$_2$ and a detailed comparison with those of other Fe-pnictides will provide more insight into the underlying electronic scattering mechanism, that is intimately coupled to the nature of superconducting paring in Fe-pnictides.

Here, we present a detailed study of the normal-state transport properties of isovalent Ru-substituted BaFe$_{2-x}$Ru$_x$As$_2$ using high-quality single crystals in a wide range of Ru substitution (0 $\leq$ $x$ $\leq$ 1.4) covering the AFM, superconducting (SC), and paramagnetic(PM) phases. In agreement with previous works,\cite{Ba122_Ru:colson:syn,Ba122_Ru:canfield:syn,Ba122_Ru:canfield:structure,Ba122_Ru:canfield:TEP} we found that Ru substitution weakens the AFM order and eventually induces the superconducting phase above relatively high doping levels of $x$ $>$ 0.4. Near the AFM phase boundary, the electrical resistivity, $\rho(T)$, shows a linear $T$-dependence, and the Hall constant, $R_H(T)$, exhibits a strong temperature dependence with a \emph{sign-change} at $T$ $\sim$ 100 K, suggesting possible non-Fermi-liquid behavior. Exploring higher substitution levels than in previous works\cite{Ba122_Ru:colson:syn,Ba122_Ru:canfield:syn,Ba122_Ru:canfield:structure,Ba122_Ru:canfield:TEP}, which enable us to close the SC dome, we found that Fermi-liquid behavior is recovered as indicated by the temperature exponent of $\rho(T)$ approaching $\sim$ 2 again. In the highly doped regime, $R_H(T)$ becomes almost temperature-independent and negative throughout the whole temperature range.
In contrast to the strong doping dependence of $\rho(T)$ and $R_H(T)$, the magnetoresistance, that is typically expected to be large and doping-sensitive for systems with multi-band structures, remains small and almost doping-independent. These observations suggest that the anomalous transport properties in the normal state of Ru-substituted BaFe$_2$As$_2$ cannot be explained by multi-band scenario. We rather attribute the anomalous transport properties to the strong anisotropy of the scattering rate in the electron FS due to the interband AFM fluctuations.

Single crystals of BaFe$_{2-x}$Ru$_x$As$_2$ were grown from FeAs and RuAs flux using similar methods as described in Refs. \onlinecite{Ba122_Ru:colson:syn,Ba122_Ru:canfield:syn}. Plate-shaped crystals with a shiny (001) surface were extracted mechanically. X-ray diffraction on single crystals revealed sharp (00$l$) peaks, confirming a successful single crystal growth. The doping level for each crystal was determined by energy dispersive X-ray spectroscopy. The in-plane resistivity and the Hall coefficient were measured using the standard 6-probe method in a Physical Property Measurement System (PPMS-14T, Quantum Design). Some of the crystals were annealed in Ar atmosphere at 650 $^{\rm o}$C for a week. Such crystals show a shaper superconducting transition than the as-grown crystals. However, the transport properties in the normal state are consistent with each other for as-grown and annealed crystals. Magnetic susceptibility measurements were done in a magnetic field of 10 Oe along the $ab$-plane using a SQUID magnetometer(MPMS, Quantum Design).

\begin{figure}
\includegraphics*[width=8.5cm, bb=55 55 540 440]{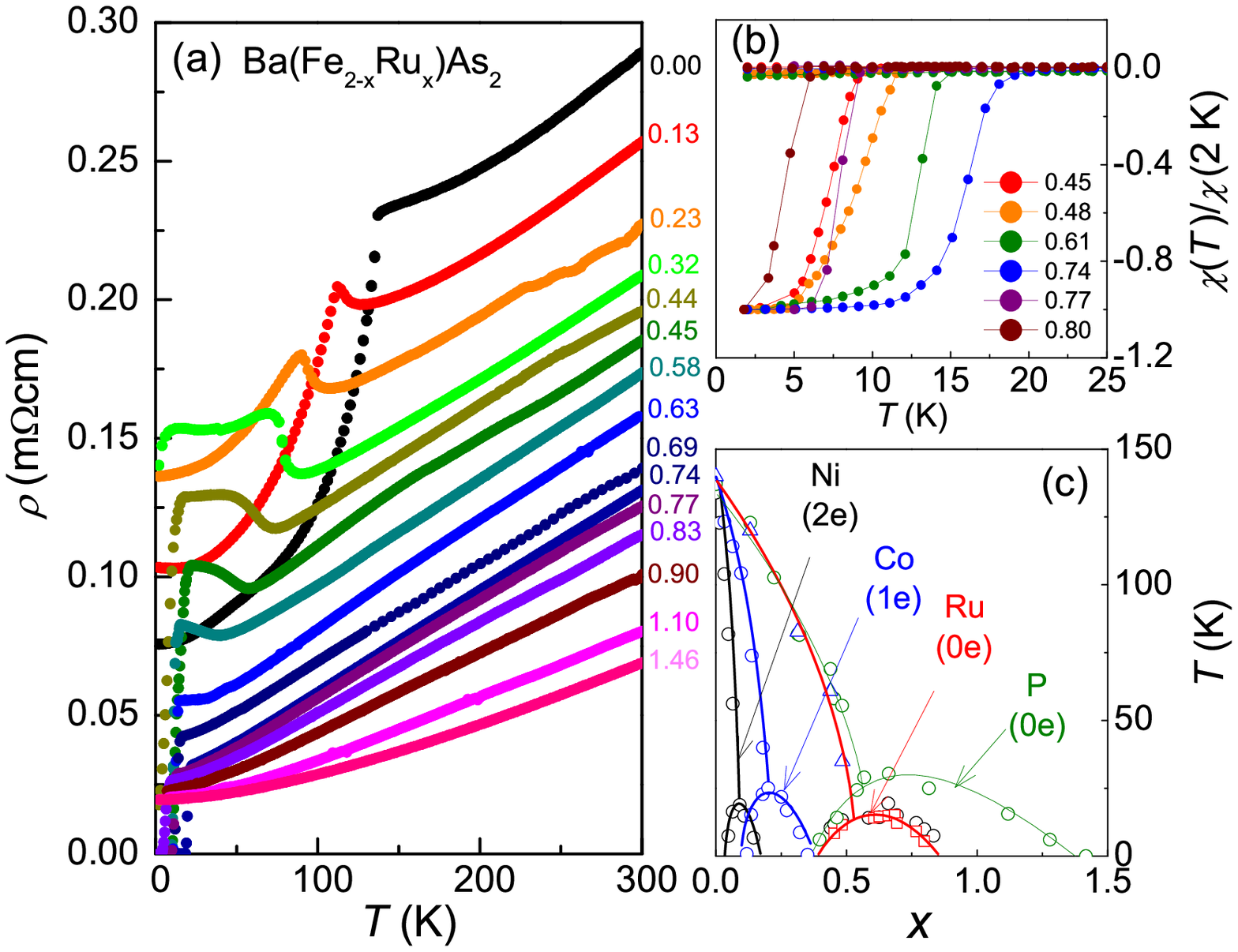}
\caption{\label{fig1}(Color online)
(a)Temperature dependence of the in-plane resistivity $\rho(T)$ of BaFe$_{2-x}$Ru$_x$As$_2$ single crystals(0 $\leq$ $x$ $\leq$ 1.5). (b) Temperature dependence of the magnetic susceptibility for the same BaFe$_{2-x}$Ru$_x$As$_2$ single crystals near $T_c$ at $H$ = 10 Oe. (c) Phase diagrams of BaFe$_{2-x}$$M_x$As$_2$ where $M$ = Ni(Ref. \onlinecite{Ba122_Ni:li:syn}), Co(Ref. \onlinecite{Ba122_Co:hhwen:hall}), and Ru (this work) as well as BaFe$_2$$As_{2-x}$P$_x$ (Ref. \onlinecite{Ba122_P:matsuda:syn}). The excess charge introduced by doping is indicated in the parentheses.}
\end{figure}

Figure 1(a) shows the in-plane resistivity $\rho(T)$ as a function of temperature. For the parent compound BaFe$_2$As$_2$, an anomaly in $\rho(T)$ is obtained at $T_{\rm AFM}$ = 137 K corresponding to the AFM transition. With Ru substitution, $\rho(T)$ has lower resistivities, and the anomaly is systematically shifted to lower temperatures, indicating the suppression of the AFM order. Superconducting transition occurs at $x$ $>$ 0.4 confirmed by the drop in $\rho(T)$ as well as by the diamagnetic signal in $\chi(T)$ (Fig.\ref{fig1}(b)). The maximum $\Tc$ $\sim$ 20 K is observed at $x$ $\approx$ 0.7 in good agreement with the previous reports\cite{Ba122_Ru:colson:syn,Ba122_Ru:canfield:syn}. Extending to higher subsitution levels, we found that the superconducting dome closes at $x$ $\sim$ 0.9, and even more metallic phases are induced. This is consistent with the paramagnetic metallic phase found in polycrystalline BaRu$_2$As$_2$.\cite{Ba122_Ru:johnston:syn}

The phase diagram of BaFe$_{2-x}$Ru$_x$As$_2$ derived from the anomalies in $\rho(T)$ and $\chi(T)$ is very similar to those of other Fe-pnictides. The dome-shaped SC phase appears as the AFM order is suppressed, suggesting the importance of the AFM fluctuations for superconductivity. Compared to the substitution with other elements, \emph{e.g.} Co and  Ni, the suppression rate of $T_{\rm AFM}$ is significantly reduced for Ru substitution, and accordingly the range of the SC dome is shifted to higher substitution levels. While charge doping shrinks or enlarges the electron- and hole-FS's inducing a significant size-mismatch between them, isovalent substitution does not change the chemical potential, thus being less effective in deteriorating the interband nesting condition. Instead of changing the sizes of the FS's, Ru substitution alters the 3D shape of the FS as suggested by first principle calculations.\cite{Ba122_Ru:singh:band} The extended 4$d$ orbitals of Ru as well as the structural distortion of the FeAs$_4$ tetrahedra modify the degree of hybridization between Fe and As orbitals. As a consequence, the hole pockets are distorted drastically with pronounced dispersion along $k_z$, while the shape of electron pockets remains almost independent of Ru substitution. Strong warping in the hole pockets, therefore, degrades the nesting condition, thus leading to suppression of the AFM order as found also in isovalent P-substituted BaFe$_2$As$_2$\cite{Ba122_P:matsuda:syn}. Comparing the FS topology for the end member, the hole FS's in BaRu$_2$As$_2$ are disconnected at the $\Gamma$ point showing 3D FS centered at $Z$-points in the BZ, while for BaFe$_2$P$_2$ the warped cylindrical FS remains unchanged. Therefore, a somewhat bigger suppression rate for $T_{\rm AFM}$ as well as a smaller extend of the SC dome for Ru substitution can be attributed to a more significant change in the FS due to the strongly-reduced lattice anisotropy with Ru substitution\cite{Ba122_Ru:colson:syn,Ba122_Ru:canfield:syn} as compared to P substitution.

\begin{figure}
\includegraphics*[width=8.5cm, bb=55 55 540 400]{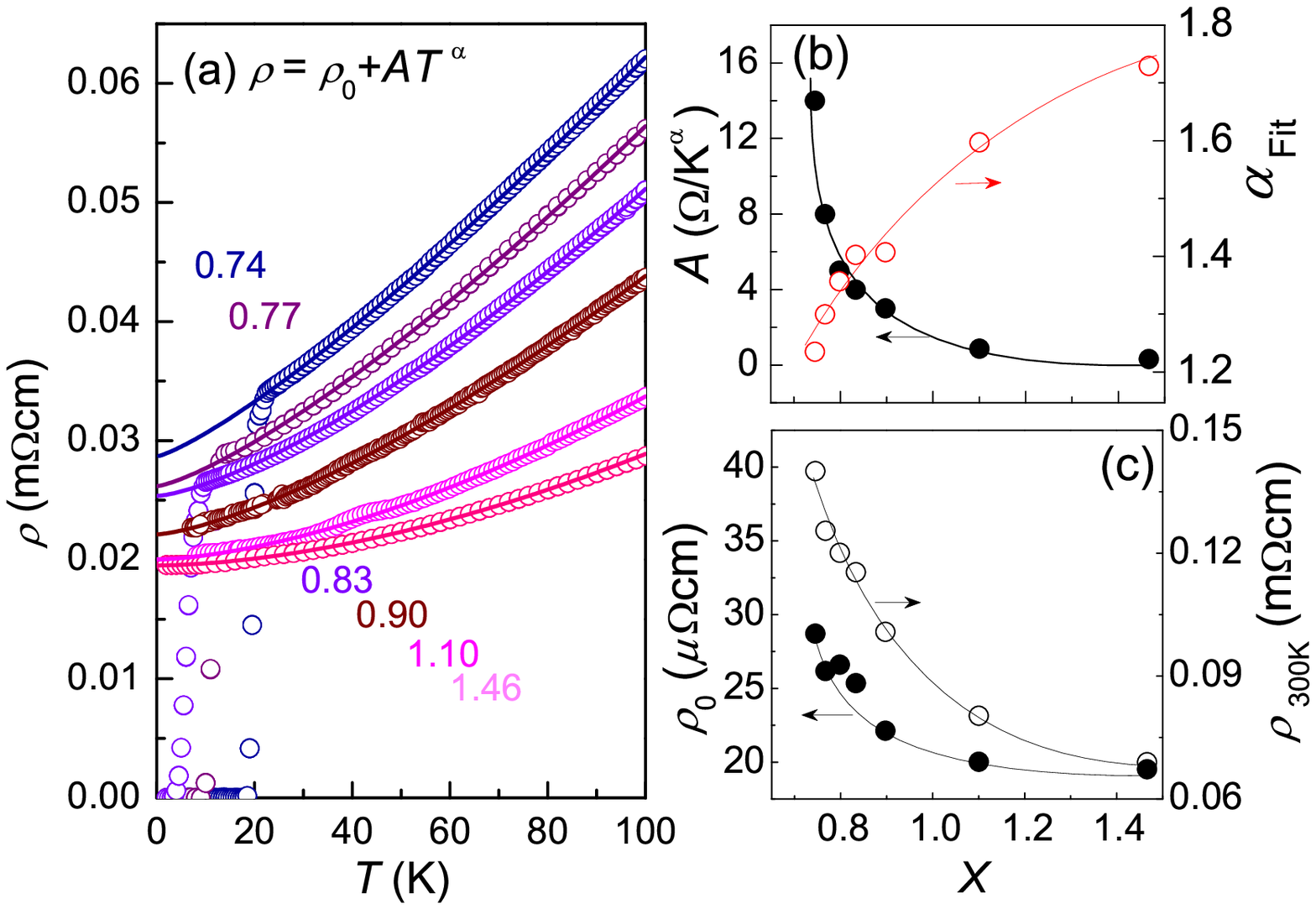}
\caption{\label{fig2}(Color online)
(a)Temperature dependence of the in-plane resistivity $\rho(T)$ for BaFe$_{2-x}$Ru$_x$As$_2$ single crystals in the normal state for $x$ $\geq$ 0.6. The solid lines are the fit of $\rho(T)=\rho_0 + A\,T^{\alpha}$ (see the text). (b) The coefficients $A$ and the exponent $\alpha$ as a function of substitution level $x$. (c) The doping dependence of the residual resistivity $\rho_0$ and $\rho$ at 300 K. The solid lines in (b)and (c) are the guide-to-eyes.}
\end{figure}

\begin{figure}
\includegraphics[width=8.5cm, bb=80 310 650 700]{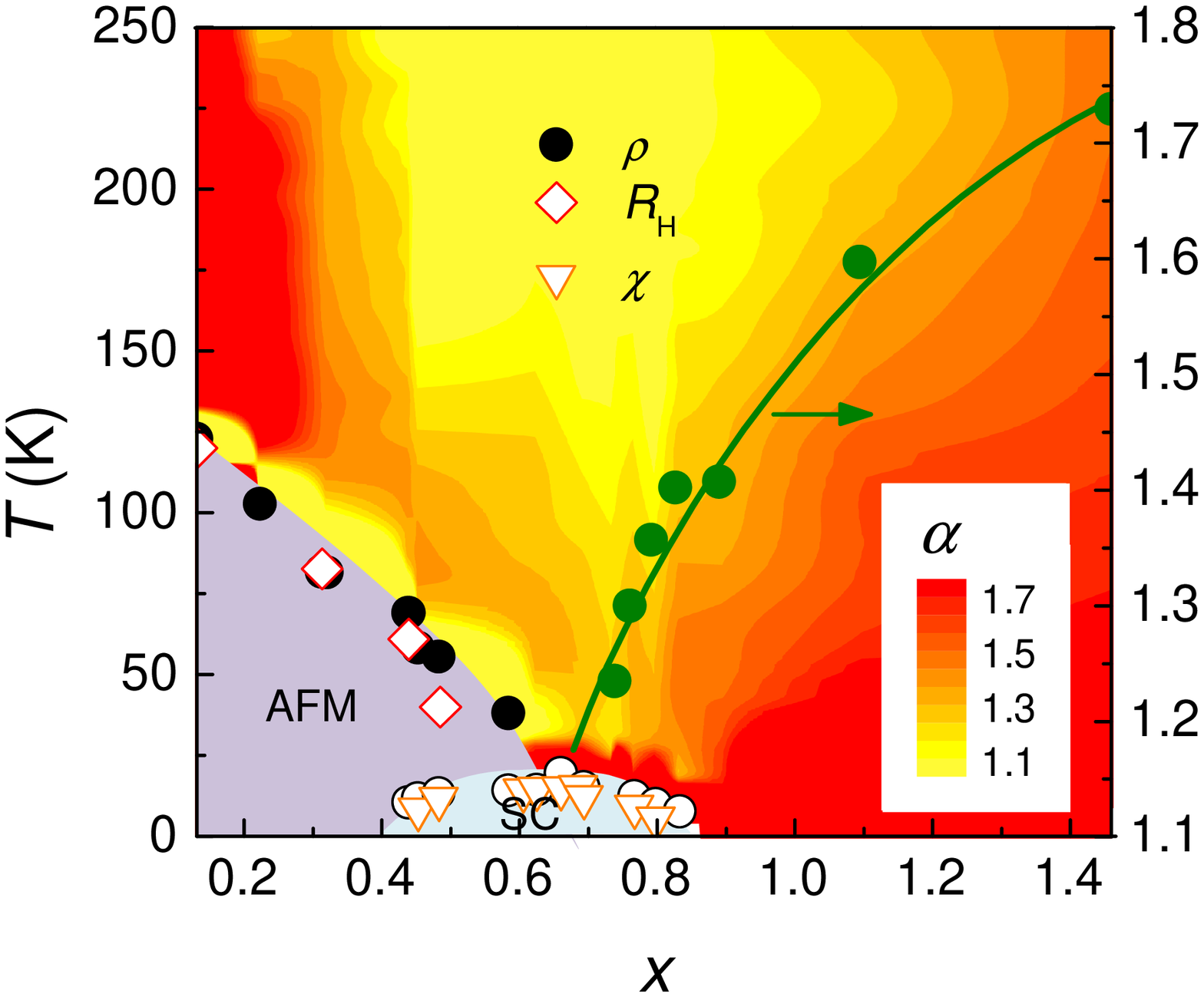}
\caption{\label{fig3}(Color online)
The phase diagram of BaFe$_{2-x}$Ru$_x$As$_2$ as a function of Ru substitution ($x$) derived from  the analysis of the resistivity ($\rho$), susceptibility($\chi$) and Hall coefficient($R_H$) data. The evolution of the temperature exponent $\alpha$ of the in-plane resistivity in the normal state is presented as a contour map. }
\end{figure}

Having established the phase diagram of BaFe$_{2-x}$Ru$_x$As$_2$, we discuss the effects of Ru substitution on the normal state electrical transport properties for a wide range of $x$. Firstly, $\rho(T)$ exhibits strong deviations from the conventional $T^2$-dependence, indicating possible non-Fermi-liquid-like behavior.
The normal state resistivity for $\Tc<T<100$ K follows nicely a power-law behavior according to $\rho(T)=\rho_0 + A\,T^{\alpha}$, where $\alpha$ is the temperature exponent and $\rho_0$ is the residual resistivity (Fig. \ref{fig2}(a)). The exponent $\alpha_{\rm fit}$ obtained by fitting the $\rho(T)$ curves for $\Tc<T<100$ K grows from $\alpha_{\rm fit}$ $\approx$ 1.2 near the AFM phase boundary at $x$ $\sim$ 0.7 to $\alpha_{\rm fit}$ $\approx$ 1.7 for $x$ = 1.46, indicating recovery of the standard $T^{2}$-behavior for $\rho(T)$ in the highly-overdoped regime. The evolution of $\alpha$ in the phase diagram is presented in a contour plot in Fig. \ref{fig3}, together with $T_{\rm AFM}(x)$ and $T_{\rm SC}(x)$. A V-shaped region with anomalous $\alpha$ is observed near the AFM phase boundary and the superconducting dome. Such a behavior is archetypical for a quantum critical point with strong AFM fluctuations and has similarly been found for various heavy fermion systems and correlated metals.

Secondly, the temperature coefficient $A$, which is related to the quasiparticle effective mass, also shows an anomalous doping dependence. Near the quantum critical point, $A$ is known to be enhanced significantly, reflecting the divergence of the effective quasi-particle mass $m^*$, as found in our system in Fig. \ref{fig2}(b), Recent de-Hass-van-Alphen experiments on BaFe$_2$As$_{2-x}$P$_x$, in fact, revealed that $m^*$ of the quasiparticles at the electron pockets is significantly enhanced near the AFM phase boundary.\cite{Ba122_P:matsuda:dHvA,Ba122_P:fisher:dHvA} These results suggest that the quasi-particle scattering is determined by the AFM fluctuations near the AFM phase boundary.

\begin{figure}
\includegraphics[width=8.0cm, bb=27 10 485 400]{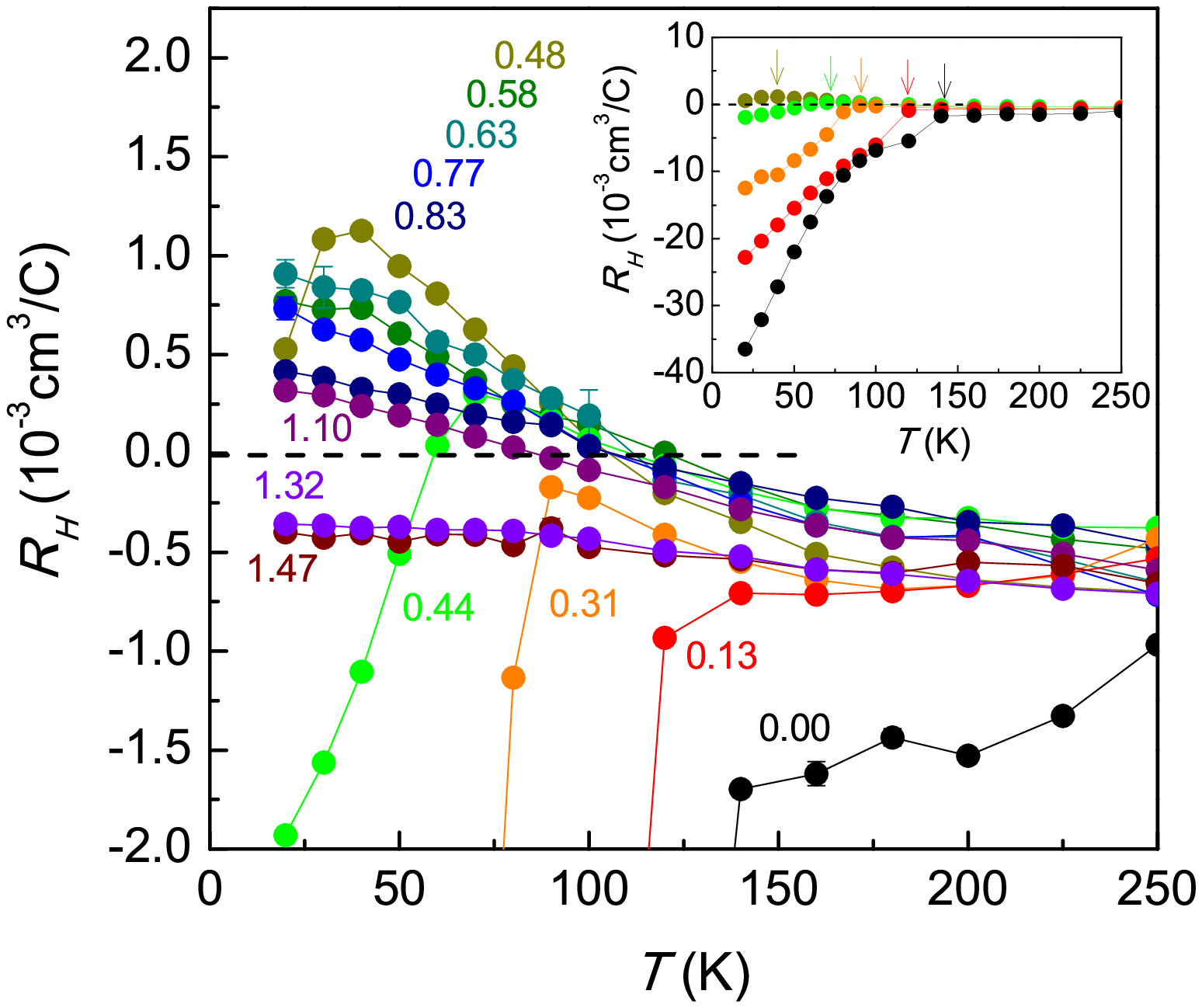}
\caption{\label{fig4}(Color online) Temperature dependence of the Hall coefficient($R_H$($T$)) for BaFe$_{2-x}$Ru$_x$As$_2$. The kink in $R_H(T)$ is due to the AFM ordering in the under-doped regime ($x$ $<$ 0.5) as indicated by arrows in the inset as well. Note that a sign change of $R_H(T)$ is observed at $T$ $\sim$ 100 K for the intermediate doping (0.3 $<$ $x$ $<$ 1.1), while $R_H(T)$ exhibits an almost temperature independent behavior for higher doping ($x$ $>$ 1.3).}
\end{figure}

Figure \ref{fig4} displays the temperature dependence of the Hall coefficient, $R_H$, for a wide range of $x$. A sudden drop of $R_H$ at $T_{\rm AFM} = 137$ K observed in BaFe$_2$As$_2$ is shifted to lower temperatures (see the inset in Fig.\ref{fig4}) with Ru substitution, consistent with the $\rho(T)$ results. In the paramagnetic state, $R_H$ is negative for undoped BaFe$_2$As$_2$ and decreases with lowering temperature, reflecting dominant contribution from the electron carriers. With Ru substitution, the initial increase with temperature ($dR_H/dT > 0$) of $R_H$ is systematically changed to a negative slope ($dR_H/dT < 0$) with a sign change at $T$ $\sim$ 100 K depending on the substitution level $x$. While $R_H$ at 250 K remains almost constant with Ru doping, the increase of $R_H$ at low temperatures becomes stronger up to $x=0.48$. Above $x > 0.5$, the negative $T$-dependence is reduced with further Ru substitution, and with $x\gtrsim1.32$, $R_H$ becomes almost temperature-independent, remaining negative over the whole temperature range. This temperature and doping dependence of $R_H$ for BaFe$_{2-x}$Ru$_x$As$_2$ is in strong contrast to Co- or P-substituted BaFe$_2$As$_2$ where $R_H$ always decreases with temperature. \cite{Ba122_Co:colson:hall,Ba122_Co:hhwen:hall,Ba122_P:matsuda:syn}

At first glance, it is tempting to attribute the strong $T$-dependence of $R_H (T)$ to multi-band transport. For compensated semi-metals with the same number of electron($n_e$) and hole carriers($n_h$), as expected for BaFe$_{2-x}$Ru$_x$As$_2$, $R_H (T)$ is described by $R_H=\frac{1}{ne}\frac{\sigma_h-\sigma_e}{\sigma_h+\sigma_e}$ with the carrier density $n=n_e=n_h\approx0.15/\rm Fe$ taken from first principle calculations and the hole(electron) conductivity $\sigma_h$($\sigma_e$). A strong $T$-dependence of $R_H$ can arise from different temperature dependences of the conductivity of the hole and electron carriers. Such an interpretation has been widely employed for Fe-pnictides systems, \emph{e.g.} Co-doped BaFe$_2$As$_2$, \cite{Ba122_Co:colson:hall,Ba122_Co:hhwen:hall} and also for Ru-substituted BaFe$_2$As$_2$\cite{Ba122_Ru:colson:syn}.

Several observations, however, suggest that this interpretation is unlikely. Firstly, in order to explain the sign-change of $R_H$ in terms of multi-band effects, one has to assume that $\sigma_e>\sigma_h$ at high temperatures, while $\sigma_e\ll\sigma_h$ at low temperatures. Since, the carrier density of $\sim 0.15/\rm Fe$ corresponds to $1/ne\approx0.98\times10^{-3}\rm cm^{3}/C$, $R_H\approx-0.5\times10^{-3}\rm m^{3}/C$ at $T=250$ K for the whole range of $x$, implies that at high temperatures $\sigma_e$ is three times larger than $\sigma_h$. At low temperatures, however, $R_H$ is enhanced up to $\approx1\times10^{-3}\rm m^{3}/C$ for $x=0.58$, which requires an almost negligible $\sigma_e$($\sigma_e\ll\sigma_h$). Such a drastic difference of the $T$-dependence of electron and hole conductivities appears to be unphysical. Secondly, the sign-change in $R_H$ is only observed in the intermediate substitution range. Consequently one has to assume that upon Ru substitution, $\sigma_h$ is enhanced quickly over $\sigma_e$ for lower doping levels, but is reduced again at higher doping level. Considering a continuous evolution of the electronic structure with Ru substitution as is also revealed by first principles calculations\cite{Ba122_Ru:singh:band}, a non-monotonous change of the ratio between $\sigma_h$ and $\sigma_e$ is rather unlikely.

\begin{figure}
\includegraphics[width=7.5cm, bb=60 65 535 535]{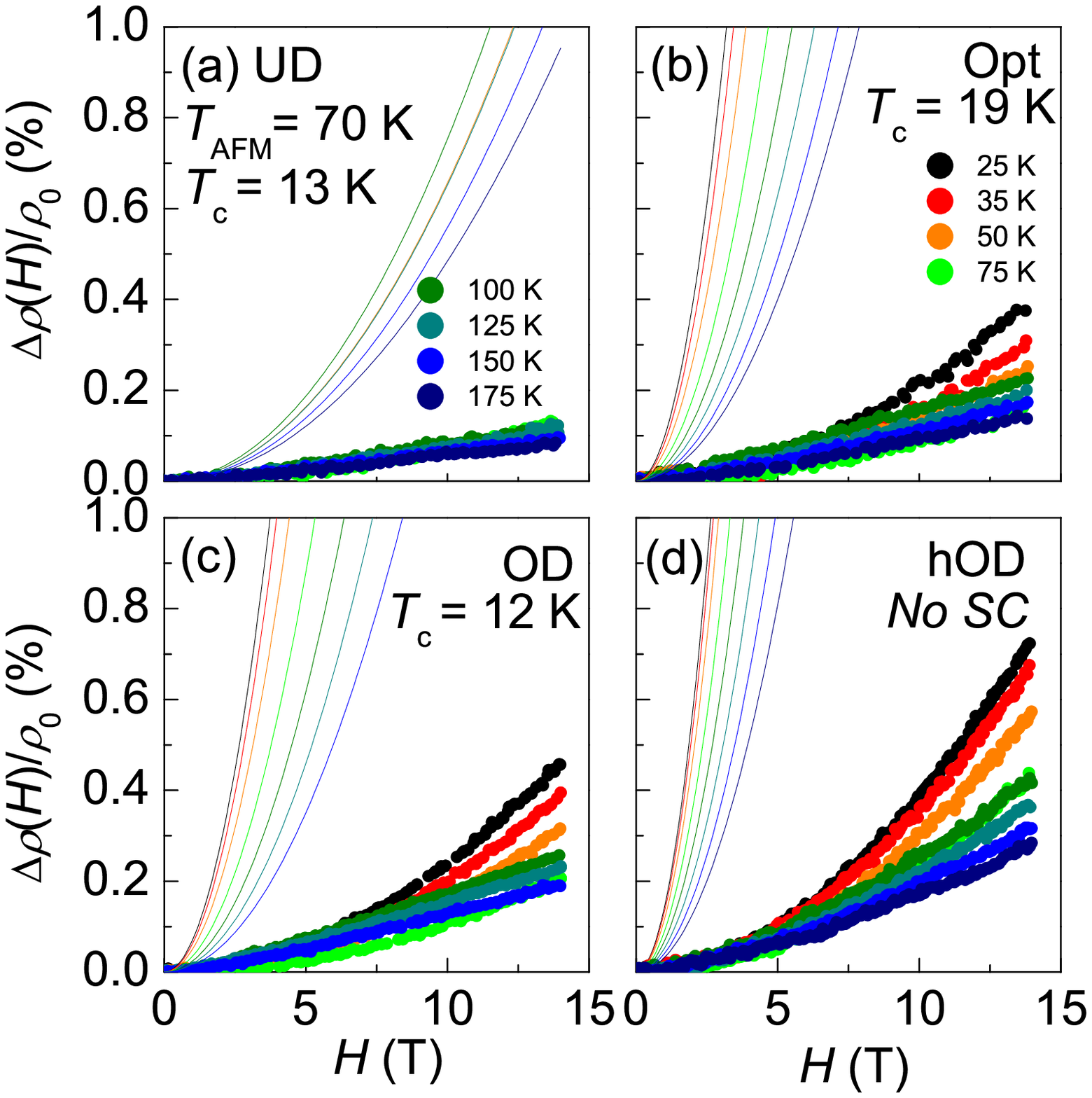}
\caption{\label{fig5}(Color online) Magnetoresistance in the paramagnetic state at various temperatures for under-doped (UD, $x$ = 0.45, $T_{\rm AFM}$ = 75 K and $T_c$ = 12 K optimally-doped (Opt, $x$=0.70 and T$_c$= 19 K), over-doped(OD, $x$=0.85 and $T_c$ = 12 K), and heavily over-doped (hOD, $x$= 1.30 without superconductivity) BaFe$_{2-x}$Ru$_x$As$_2$. Note that for under-doped samples the MR curves above $T_{\rm AFM}$ are presented. The solid lines indicate the calculated magnetoresistance based on the two-band model (see the text).}
\end{figure}
Further experimental evidence against multi-band effects is gained from the magnetoresistance (MR) results. In a multi-band system, the MR is described by the following expression, assuming average conductivity $\sigma_i$, cyclotron frequency $\omega_{ci}$, and relaxation time $\tau_i$ for each band,
\begin{equation}
\frac{\Delta \rho}{\rho_0} \approx \frac{1}{2} \frac{\Sigma_i \Sigma_{j\neq i} \sigma_i \sigma_j (\omega_{ci} \tau_i - \omega_{cj} \tau_j)^2 }{(\Sigma_i \sigma_i)^2}.
\end{equation}
,where $\omega_{ci}$ terms have opposite signs for electron and hole bands. In this case, the $(\omega_{ci} \tau_i - \omega_{cj} \tau_j)^2$ term and also the resulting MR become larger because the $\omega_{ci} \tau_i$ terms add up. For example, MgB$_2$, one of the well-known multi-band systems, has electron($\pi$) and hole($\sigma$) FS's and exhibits an almost 100 $\%$ MR at low temperatures.\cite{MgB2:wen:MR} Therefore, if  multi-band effects govern the normal state transport properties of BaFe$_{2-x}$Ru$_x$As$_2$, one would expect a large MR since electron and hole FS's coexist. Figure 5 shows the MR of BaFe$_{2-x}$Ru$_x$As$_2$ samples in the PM state. Across the whole doping range,
the MR remains less than 1 $\%$, and in particular, it does not exhibit any significant doping dependence.

For a quantitative comparison between the multi-band scenario and the experimental observation, we calculated the MR based on the two-band model using Eq. (1) with the constraints that the experimentally measured $1/\rho(T)$ = $\sigma_e$+$\sigma_h$ and $R_H(T)$ = (1/$ne$)($\sigma_h$-$\sigma_e$)/($\sigma_h$+$\sigma_e$). The charge carrier density $n$ = $n_e$ = $n_h$ = $~0.15/\rm Fe$ is assumed to be constant with isovalent Ru substitution. As shown in Fig. \ref{fig5}, the calculated MR is almost two orders of magnitude larger than experimentally observed. We tested several values of $n_{e,h}$, considering their possible doping dependence as proposed in Ref.\onlinecite{Ba122_Ru:colson:syn}, but the calculated MR never comes close to the experimental values. The main reason for this large discrepancy is that the two-band model predicts a large MR as long as the electron and hole carriers coexist and their mobilities are comparable. Thus even in more realistic multi-band models including more than two FS's, the discrepancy remains qualitatively the same. These observations clearly demonstrate that there is a strong disparity between electron and hole carriers, and the transport properties in the normal state of BaFe$_{2-x}$Ru$_x$As$_2$ is dominated by one type of carriers. In fact, there is growing experimental evidence showing that the carriers in the \emph{electron} FS determine the transport properties in the normal states.
For example, recent Raman scattering experiments on Co-substituted BaFe$_2$As$_2$\cite{Ba122_Co:hackl:Raman} and quantum oscillations in isovalent P-substituted BaFe$_2$As$_2$\cite{Ba122_P:matsuda:dHvA,Ba122_P:fisher:dHvA} indeed revealed a much higher mobility in the electron FS.

\begin{figure}
\includegraphics[width=7.5cm, bb=40 70 500 780]{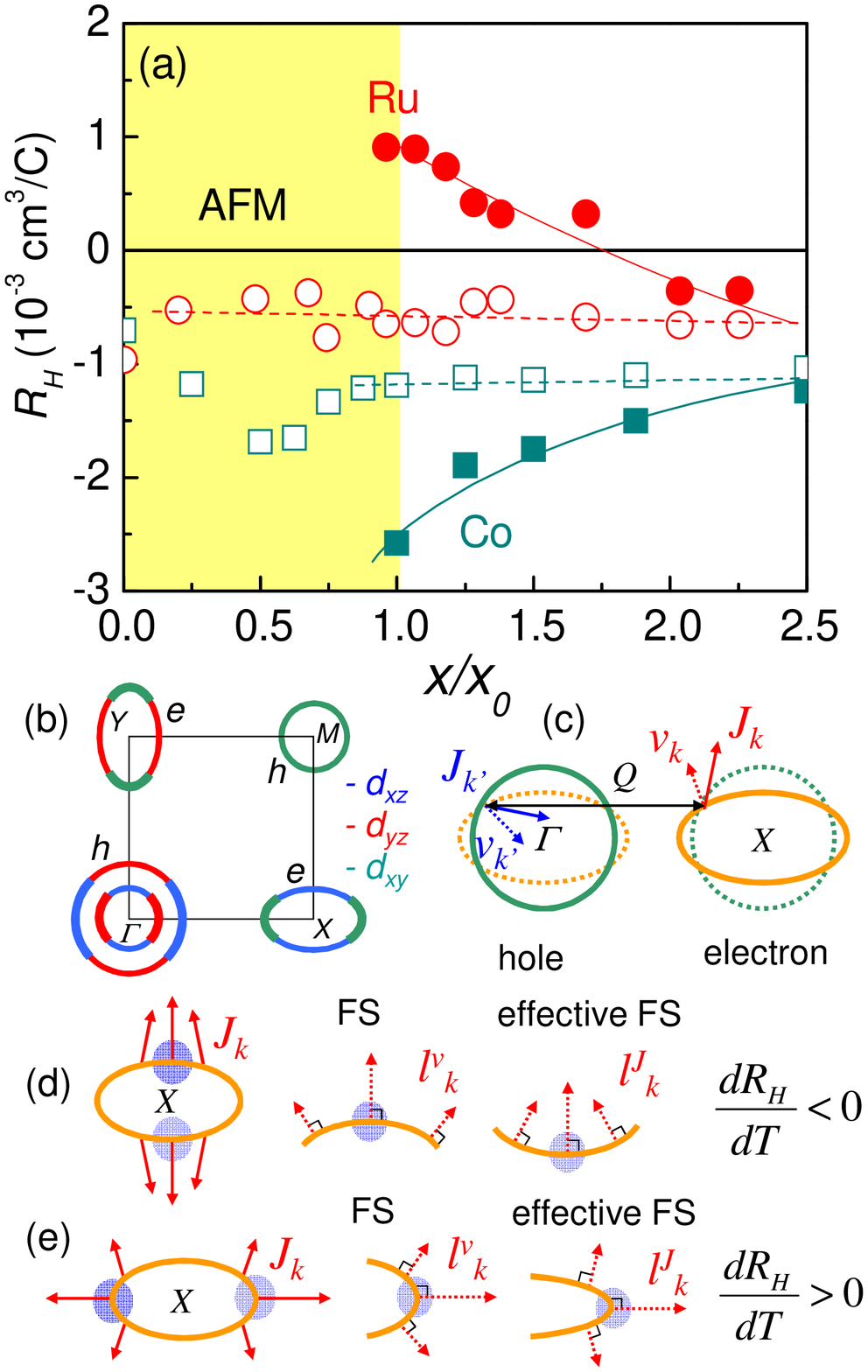}
\caption{\label{fig6}(Color online) (a) The Hall coefficient ($R_H$) of BaFe$_{2-x}$Ru$_x$As$_2$ and BaFe$_{2-x}$Co$_x$As$_2$ (Ref. \onlinecite{Ba122_Co:hhwen:hall}) at $T$ = 20 K (solid) and 250 K (open) as a function of the normalized doping level ($x$/$x_0$) where $x_0$ corresponds to the optimal doping. (b) Typical FS of 122 Fe-pnictides, showing different Fe orbital characters. (c) Two FS model with a hole(electron) pocket at the $\Gamma$($X$) point in the unfolded BZ. Due to strong AFM scattering with $Q$ = ($\pi$, 0) (dashed line), that couples quasi-particles in hole ($k$) and electron pockets ($k+Q$), the current flow $J_k$ (solid arrow) is tilted from the Fermi velocity $v_k$ (dashed arrow). (d)(e) The FS (normal to $v_k$ $\sim$ $l^v_k$) and the "effective" FS (normal to $J_k$ $\sim$ $l^v_J$) around a cold spot (indicated with blue circles). Here $l^v_k$ and $l^v_J$ are the corresponding mean-free-paths. Note that the curvature of the "effective" FS is opposite (enhanced) when the cold spot is located on the short(long) axis of elliptical electron FS in (d)((e)), which leads to a strong $T$-dependence of $R_H(T)$ with a different sign of $dR_H/dT$.}
\end{figure}

Let us now consider the alternative scenario for the unconventional doping dependence of $R_H(T)$. In Fig. \ref{fig6}(a), we plot $R_H$ at low (20 K) and high temperatures (250 K) for Ru-doped and Co-doped BaFe$_2$As$_2$ \cite{Ba122_Co:hhwen:hall} as a function of the normalized doping level $x$/$x_0$, where $x_0$ corresponds to the optimal doping indicated by maximum $T_c$. We focus only on the transport properties in the PM state, thus the significantly large values of $R_H$ found in the low temperature AFM state in the low doping regime are not considered. At high temperatures, $R_H$ remains negative without significant doping dependence, suggesting the dominant role of electron carriers in the transport properties. At low temperatures, however, $R_H$ exhibits a strong doping dependence for both cases of Ru- and Co- substitution. As the system approaches the AFM phase ($x/x_0$ $\sim$ 1), $R_H$ at 20 K strongly deviates from that at 250 K indicating a strong $T$-dependence, while such a deviation is almost completely suppressed when the system is placed far from the AFM phase ($x/x_0$ $\sim$ 2.5). Similar strong $T$-dependence of $R_H$ has been also observed in K- or P-substituted BaFe$_2$As$_2$.\cite{Ba122_K:yuan:hall,Ba122_P:matsuda:syn} This dependence implies that the proximity to the AFM phase is essential for the unconventional behavior of $R_H(T)$.

Despite of the similarity in the vicinity of the AFM phase, the slope of the $T$-dependence of $R_H(T)$ is opposite for Ru- and Co-substitution; Ru substitution leads to a negative slope  ($dR_H/dT$ $<$ 0) as shown in Fig. \ref{fig4} while Co doping results in a positive slope ($dR_H/dT$ $>$ 0)\cite{Ba122_Co:hhwen:hall}. Although systematic studies are not available at present, similar tendencies have been found in other 122 pnictides, such as K- or P-substituted BaFe$_2$As$_2$. \cite{Ba122_K:yuan:hall,Ba122_P:matsuda:syn} Thus according to their $T$-dependence of $R_H$ there appear to be two groups of 122 systems; Ru- and K-substituted BaFe$_2$As$_2$ belong to a group with $dR_H/dT$ $<$ 0, while Co- and P-substituted BaFe$_2$As$_2$ belong to a group with $dR_H/dT$ $>$ 0. Therefore the sign of the slope in $R_H$ reflects the detailed change in electronic structures with doping for BaFe$_2$As$_2$.

Similar anomalous doping- and compound-dependence of $R_H(T)$ as well as $\rho(T)$ have, in fact, been observed in high-$T_c$ cuprates and heavy fermion compounds. For example, La$_{2-x}$Sr$_x$CuO$_4$ and Nd$_{2-x}$Ce$_x$CuO$_4$ exhibit strongly $T$-dependent $R_H(T)$ as well as a linear-$T$ dependence of $\rho(T)$ near the AFM or pseudogap phase boundary\cite{cuprate:sato:hall,cuprate:hwang:hall,cuprate:greene:hall}. The magnitude of $R_H$ is enhanced in both La$_{2-x}$Sr$_x$CuO$_4$ and Nd$_{2-x}$Ce$_x$CuO$_4$ at low temperatures, but their temperature dependence implies different sign of $dR_H/dT$ : negative (positive) for La$_{2-x}$Sr$_x$CuO$_4$ (Nd$_{2-x}$Ce$_x$CuO$_4$). A similar behavior of $R_H(T)$ has also been reported for Ce$M$In$_5$ ($M$ = transition metal).\cite{HF:kontani:hall,HF:nakajima:hall}

In these systems, there has been growing evidence that anisotropic scattering at the FS is essential for the anomalous transport properties commonly observed in the vicinity of the AFM phase.\cite{AFmetal:kontani:transport} In a nearly AFM metal, electron-electron scattering is dominant at a certain $Q$ vector. This induces a strong disparity in scattering times within the same FS, producing the so-called hot(cold) spots with a larger(smaller) scattering rate. 
Furthermore, the strong AFM scattering process in nearly AFM metals changes the direction of electron motion. Considering the current vertex correction due to back-flow current, the current $J_k$ is not parallel to the quasi-particle velocity $v_k$ but parallel to $\approx$ $v_k$+$v_{k+Q}$. This implies that 
$R_H$ is not measuring the curvature of the original FS (normal to $v_k$), but probing the curvature of an "effective" FS that is normal to $J_k$. Since such an anomalous $k$-dependence of $J_k$ and its anisotropic scattering becomes more pronounced as the system approaches to the AFM phase boundary, the magnitude of $R_H$ is strongly enhanced at low temperatures or at low doping level.

In the case of Fe-pnictides, the AFM ordering is related to the interband nesting via $Q$ = ($\pi$, 0) or (0, $\pi$) between the circular hole pockets at the $\Gamma$ and $M$ points and elliptical electron pockets at the $X$ or $Y$ points in the unfolded BZ as shown in Fig. \ref{fig6}(b). Considering the vertex correction due to the dominant AFM scattering, $J_k$ can significantly deviate from being normal to the original FS. In a simplified two FS's model with a circular hole and an elliptical electron pocket (Fig. \ref{fig6}(c)), an excited electron at $k$ in the electron pocket is scattered to $k+Q$ in the hole pocket due to the AFM scattering with $Q$ = ($\pi$, 0), thus leading to $J_k$ $\propto$ $v_k$+$v_{k+Q}$. Note that similar deviations of $J_k$ occur in the other electron pockets at the $Y$ point in the unfolded BZ, thus the effect is $additive$ for all the electron pockets. On the other hand, for the hole pocket, the combined effects of ($\pi$, 0) and (0, $\pi$) scattering with electron pockets at $X$ and $Y$ points cancel each other, reducing the back-flow effects on the hole pocket. The Back-flow effects lead to an "effective" FS, which can be quite different from the original FS of the electron pocket, as shown in Fig. \ref{fig6}(d) and \ref{fig6}(e). In particular, along the short axis of the elliptical electron FS, the curvature is opposite, which results in a positive contribution to $R_H$ (Fig. \ref{fig6}(d)). In contrast, along the long axis in the elliptical electron FS, the curvature becomes larger, thus generating a negative contribution to $R_H$ (Fig. \ref{fig6} (e)). Since the back-flow effect is proportional to the coherence length $\xi_{AFM}$ of the AFM fluctuations, this effect becomes more pronounced at lower temperatures, thus leading to the observed strong $T$-dependence of $R_H(T)$. The sign of d$R_H$/d$T$ is determined by that section of the electron FS which contributes dominantly to the transport properties, in other words, where the cold spot is located.

Recent theoretical studies, in fact, support the hot/cold spot structure in the electron pocket for Fe-pnictides.\cite{pnictide:kontani:transport,pnictidel:kemper:transport} In addition, the multi-orbital structure of each Fermi pocket also affects the anisotropy of the scattering rate. As shown in Fig. \ref{fig6}(b) both electron and hole pockets consist of the FS sections with $d_{xz}$/$d_{yz}$ and $d_{x^2-y^2}$ orbital character. The interband scattering rate is sensitive not only to the relative size between hole and electron pockets, but also to the relative weight of the orbital character. Therefore the temperature dependence of $R_H(T)$ reflects the anisotropic nature of the interband AFM scattering. The opposite $T$-dependence of $R_H(T)$ implies different hot/cold structure in the electron pockets of Co- and Ru-substituted BaFe$_2$As$_2$. For undoped BaFe$_2$As$_2$ with $dR_H/dT$ $<$ 0, the cold spot is on the FS section near the long axis of the electron pockets, which remains the same for Co substitution. In contrast, Ru substitution induces a relocation of the cold spots, $i.e.$ cold spots are placed near the short axis of the electron pockets, resulting in $dR_H/dT$ $>$ 0.

These observations indicate that the intra-pocket anisotropy $i.e.$ hot/cold spot structure as well as the back-flow effects due to AFM scattering in the electron pocket are essential to understand the strong temperature-/doping-/compound-dependence of the transport properties, in particular, of $R_H(T)$. Theoretically, different locations of the cold spot were predicted, depending on the dominant scattering mechanisms. Onari and Kontani suggested that with dominant orbital fluctuations, the cold spots are located in the FS section with $d_{xz}$/$d_{yz}$ orbital character near the short axis.\cite{pnictide:kontani:transport} On the other hand, Kemper \emph{et al.} showed that the FS section with $d_{x^2-y^2}$ orbital character near the long axis becomes the cold spot since Hund coupling favors intra-orbital rather than inter-orbital scattering.\cite{pnictidel:kemper:transport} A detailed comparison of the orbital character and the relative size of the FS's for Co- and Ru-doped BaFe$_2$As$_2$ will provide better understanding on the dominant scattering mechanism in Fe-pnictides.

In summary, we have carried out a detailed investigation of the magneto-transport properties in the normal state of isovalent Ru-substituted BaFe$_2$As$_2$ using high-quality single crystals in a wide range of Ru doping (0 $\leq$ $x$ $\leq$ 1.4). Our results show that the intrapocket scattering anisotropy in the electron pockets and the back-flow effects have to be taken into account to understand the non-Fermi-liquid-like features, in particular, a significant $T$-dependence of $R_H(T)$ with a sign-change. Ru-substituted BaFe$_2$As$_2$ exhibits an opposite $T$-dependence of $R_H(T)$ as compared to the undoped mother compound BaFe$_2$As$_2$, which is also in contrast to Co- and P-substituted BaFe$_2$As$_2$. This suggests that the locations of hot/cold spots are reversed with Ru substitution, while they remain the same with Co- or P- substitution. Recently, Qiu \emph{et al.} proposed a nodal superconducting gap for Ru-substituted BaFe$_2$As$_2$ based on the thermal conductivity measurements.\cite{Ba122_Ru:Li:thermal} This result indicates a strong anisotropy of superconducting pairing channel, which is consistent with anisotropic AFM scattering found in the normal-state transport properties. Locating the node in the superconducting gap and clarifying its relation to the hot/cold structure in Ru-doped BaFe$_2$As$_2$ is highly desirable.

\acknowledgments The authors acknowledge J. P. Koo, K. B. Lee, Kee Hoon Kim and H. Kontani for experimental advices and useful discussion. This work was supported by Leading Foreign Research Institute Recruitment Program(2010-00471), Basic Science Research Programs (2010-0005669) through the National Research Foundation of Korea(NRF), and the POSTECH Basic Science Research Institute Grant.

\end{document}